\documentclass[prd,aps,preprint,nofootinbib]{revtex4}
\usepackage{epsfig}
\usepackage{amsmath}
\usepackage{amssymb}

\begin{document}

\preprint{BNL-NT-07/35} \preprint{RBRC-687}

\vspace*{2cm}
\title{Hadronic Dijet Imbalance and Transverse-Momentum Dependent
Parton Distributions}

\author{Werner Vogelsang}
\email{vogelsan@quark.phy.bnl.gov} \affiliation{Physics
Department, Brookhaven National Laboratory, Upton, NY 11973}
\author{Feng Yuan}
\email{fyuan@quark.phy.bnl.gov}
\affiliation{RIKEN BNL Research Center, Building 510A, Brookhaven
National Laboratory, Upton, NY 11973 \\ \\}
\begin{abstract}
We compare several recent theoretical studies of the single
transverse spin asymmetry in dijet-correlations at hadron colliders.
We show that the results of these studies are all consistent. To establish
this, we investigate in particular the two-gluon exchange contributions
to the relevant initial and final state interactions in the context of a
simplifying model. Overall, the results confirm that the dijet imbalance
obeys at best a non-standard or ``generalized'' transverse-momentum-dependent
factorization.
\end{abstract}
\maketitle

\newcommand{\be}{\begin{equation}}
\newcommand{\ee}{\end{equation}}
\newcommand{\ben}{\[}
\newcommand{\een}{\]}
\newcommand{\beqn}{\begin{eqnarray}}
\newcommand{\eeqn}{\end{eqnarray}}
\newcommand{\Tr}{{\rm Tr} }

{\bf 1. Introduction.} Recently, there has been tremendous
interest in the ``imbalance'' of two jets that are produced nearly
back-to-back in azimuthal angle in a hadronic reaction. Especially
the single transverse spin asymmetry (SSA) for this process has
received a lot of attention in theoretical work
\cite{BoeVog03,mulders,mulders1,VogYua05,dijet-cor1,qvy-short,qvy-long,collinsqiu,RT},
while experimental studies have begun at the
Relativistic-Heavy-Ion-Collider (RHIC) at Brookhaven National
Laboratory (BNL) \cite{star-dijet1}. In Ref.~\cite{BoeVog03}, it
was proposed to use dijet-correlations at RHIC to learn about
the transverse-spin and transverse
momentum dependent (TMD) Sivers functions~\cite{Siv90}. A
transversely polarized proton (with momentum $P_A$ and
polarization vector $S_\perp$) is scattered off an unpolarized proton
with momentum $P_B$, producing two jets with momenta $P_1$ and
$P_2$:
\begin{equation}
A(P_A,S_\perp)+B(P_B)\rightarrow J_1(P_1)+J_2(P_2)+X\ .
\end{equation}
For convenience, two additional momentum vectors, $P=(P_1+P_2)/2$
and $q=P_1-P_2$, are introduced. The transverse momentum components
of $P$ and $q$ are relevant for our discussions: $P_\perp$ is the
average of the transverse momenta of the two jets and $q_\perp$
represents their imbalance. The most interesting kinematic region
arises when the imbalance between the two jets is small:
$q_\perp\ll P_\perp$, so that there are two very separate momentum
scales in the problem.

Much of the very recent theoretical work has centered on the question if
cross sections for the process may be factorized in terms of TMD
functions in this kinematic regime
\cite{mulders,mulders1,qvy-short,qvy-long,collinsqiu}.
This question has been addressed by three groups, in the following
ways:
\begin{description}
\item[(1)] in Refs.~\cite{mulders,mulders1}, the gauge-link structure of the
TMD parton correlators appearing in the dijet process was investigated.
It had been
known beforehand \cite{{BroHwaSch02},{Col02},{BelJiYua02},{Boer:2003cm}}
that Sivers-type SSAs in semi-inclusive deep inelastic scattering
(SIDIS) or the Drell-Yan process may be generated by final-state
or initial-state interactions, respectively, of the involved partons from
the polarized hadron. These interactions sum to a simple gauge
link in both cases, which carries the color charge of that parton,
albeit -- famously --
with an opposite sign between the two processes. As was discussed
in~\cite{mulders,mulders1}, the SSA for dijet (or dihadron) production is associated
with both initial-state {\it and} final-state interactions of the partons
involved in the hard-scattering. It was found that if all interactions are
summed up, the resulting gauge link for the TMD parton distributions takes
a much more complicated form. It does not involve just the color charge
of the relevant parton, but has in general knowledge about the full
hard-scattering process and its color structure. As such, different
correlators were found to appear in different partonic channels, even if
the parton type entering from the polarized proton is the same. This
observation makes the non-universality of the TMD parton distributions
much more dramatic than previously indicated by their sign difference
between the SIDIS and Drell-Yan processes. It was also found
in~\cite{mulders,mulders1} that a certain weighted moment of the spin-dependent
cross section reduces the expression to a more standard form
akin to collinear factorization, involving the twist-three matrix elements
of~\cite{et,qiusterman} and special hard-scattering factors.
\item[(2)] in Refs.~\cite{qvy-short,qvy-long}, a different approach was taken.
The SSA for the dijet imbalance was examined at first order in perturbation
theory, starting from collinear factorization in the intermediate transverse
momentum region $\Lambda_{\rm QCD}\ll q_\perp\ll P_\perp$, and carefully
examining the one-gluon radiation contributions. It was found that, {\it at
this order}, the cross sections do factorize into TMD parton distributions
that follow their definitions in either the SIDIS or the Drell-Yan process,
while all the initial-state and final-state interaction effects can
be absorbed into the hard-scattering factors. These leading-order hard
factors can be calculated either in a model-inspired fashion or in a
partonic scattering picture~\cite{qvy-long}. They coincide with the
ones found in~\cite{mulders,mulders1} for the weighted 
spin-dependent cross section
mentioned above.
\item[(3)] Ref.~\cite{collinsqiu} stresses that the non-trivial gauge link
structure found in~\cite{mulders,mulders1} implies that a ``standard'' 
factorization
in terms of TMD parton distributions cannot hold for this process.
A particularly transparent example is given in terms of a simple abelian
model in which the two scattering partons are assumed to have different and
unrelated charges, $g_1$ and $g_2$. By a first-order calculation
it is suggested that indeed the gauge link to be associated with the
parton of charge $g_1$ will in general depend on the charge $g_2$
as well, in violation of standard factorization. From a phenomenological
point of view, TMD correlators extracted from dijet-correlations would
have no connection with those from the SIDIS and Drell-Yan processes,
because their definitions are different. It is argued that if a standard
definition of parton distributions is kept, there will be uncancelled
singularities at higher order of perturbation theory. This happens for
the spin-dependent as well as the spin-averaged cross section.
\end{description}

The goal of the present paper is to put the results of
Refs.~\cite{mulders,mulders1,qvy-short,qvy-long,collinsqiu} into
context\footnote{We will from now on use for simplicity the
labels ``(1),(2),(3)'' to refer to the
papers~\cite{mulders,mulders1},~\cite{qvy-short,qvy-long}, 
and~\cite{collinsqiu},
respectively, following the list given above.}.
It is clear from the above that there is not necessarily
a contradiction among the results. (1) and (3) agree in their
assessment that universality, and hence standard factorization,
of the TMD distributions is broken for this process. (2) is
based on a first-order calculation, and it is conceivable that
deviations from the factorized structure found
in Refs.~\cite{qvy-short,qvy-long} become apparent only at higher
orders. In order to shed more light on this, we perform the
following studies: first, we compare the results in (1) and (2)
by expanding the gauge links given in~\cite{mulders,mulders1} to first order
in the strong coupling. This should lead to the results of (2),
if there is mutual consistency. Furthermore, we will consider
the two-gluon contributions to the initial and final-state
interactions. This will unambiguously show if the gauge link
exponentials have the complicated and process-dependent
form presented in~\cite{mulders,mulders1} and~\cite{collinsqiu}, which
would demonstrate that indeed standard factorization breaks down
beyond the order considered in (2). It will also address
the consistency between (1) and (3). For simplicity, we will
perform the calculation in the context of the model considered
in Ref.~\cite{mulders,collinsqiu}, and only for one underlying partonic
process, $qq'\to qq'$.

{\bf 2. First-order expansion of the gauge-links in 
Refs.~\cite{mulders,mulders1}.}

Ref.~\cite{mulders,mulders1} finds the following gauge link in the
correlator for the partonic channel $qq'\to qq'$ contributing to the 
SSA for the dijet imbalance:
\begin{equation}
{\cal
U}_{qq'}=\frac{1}{N_c^2-1}\left[\left(N_c^2+1\right)\frac{{\rm
Tr}\left({\cal U}^{[\Box]}\right)}{N_c}{\cal U}^{[+]}-2{\cal
U}^{[\Box]}{\cal U}^{[+]}\right] \ ,
\end{equation}
which is normalized to the correlator for unpolarized scattering. 
Replacing the factor $1/(N_c^2-1)$ with
$1/(4N_c^2)$ will recover the normalization used in
\cite{qvy-short,{qvy-long}}. In the first-order perturbative
expansion, we will find the following contributions from the
various gauge links in the above equation,
\begin{eqnarray}
{\cal U}^{[+]}&\to& (-ig)\frac{1}{-k^++i\epsilon}\ ,\nonumber\\
{\cal U}^{[\Box]}&\to&
(-ig)\left[\frac{1}{-k^++i\epsilon}+\frac{1}{k^++i\epsilon}\right]
\ . \label{udef}
\end{eqnarray}
Here we have suppressed the color matrices. Since the latter are
traceless, one has ${\rm Tr}\left({\cal U}^{[\Box]}\right)=1$ 
to this order. Using these expansions, we get the following 
first-order expansion of the gauge link ${\cal U}_{qq'}$:
\begin{equation}
{\cal U}_{qq'}\to
(-ig)\frac{1}{N_c^2-1}\left[\left(N_c^2-3\right)\frac{1}{-k^++i\epsilon}+(-2)
\frac{1}{k^++i\epsilon}\right] \ .
\end{equation}
Comparing with the results in \cite{qvy-short,{qvy-long}}, we find that
the first term corresponds to the sum of the two final-state
interaction factors of \cite{qvy-short,{qvy-long}}, and the second 
term to the initial-state interaction factor there. More precisely, the
coefficient of $(-ig)/(-k^++i\epsilon)$ coincides with the factor
$(C_{F1}+C_{F2})/C_u$ found in column (1) of Table~I of \cite{qvy-long}, 
while the coefficient of $(-ig)/(k^++i\epsilon)$ is identical to
$C_I/C_u$. This comparison can be extended to the diagrams for the
other partonic channels, and agreement between the results 
of \cite{mulders,mulders1} and \cite{qvy-short,{qvy-long}} is found in each case.
In addition to Eq.~(\ref{udef}), one then also needs 
${\cal U}^{[-]}\to (ig)/(k^+ +i\epsilon)$. 

{\bf 3. Two-gluon exchange contribution.}

In this section we study the two-gluon exchange contribution to
the initial and final-state interaction effects in the dijet
imbalance. As we stated earlier, we will for simplicity use the abelian
model of~\cite{mulders,collinsqiu} in which the two scattering partons are
assumed to have different and unrelated charges, $g_1$ and $g_2$.
We also follow~\cite{collinsqiu} to consider only the case of
an underlying $qq'\to qq'$ hard process; however, we take the
$q$ and $q'$ as fermions and not as scalars as in \cite{collinsqiu}.

To set the stage, we write the differential cross sections for the
dijet-correlation in this model in the factorized forms {\it assumed}
in~\cite{qvy-short,qvy-long}:
\begin{eqnarray}
\frac{d\Delta\sigma(S_\perp)}
     {dy_1dy_2dP_\perp^2d^2\vec{q}_\perp}
&=& \frac{\epsilon^{\alpha\beta}S_\perp^\alpha q_\perp^\beta}
     {\vec{q}^2_\perp}
\int d^2k_{1\perp}d^2k_{2\perp}d^2\lambda_\perp
\nonumber \\
&&\times \frac{\vec{k}_{1\perp}\cdot \vec{q}_\perp}{M_P}\,
   x_a\, q_{Ta}^{\rm SIDIS}(x_a,k_{1\perp})\,
   x_b\, q_b^{\rm SIDIS}(x_b,k_{2\perp})
\label{e4}\\
&&\times \left[S_{qq'\to qq'}(\lambda_\perp)\,
      H_{qq'\to qq'}^{\rm Sivers}(P_\perp^2)\right]_c\,
\delta^{(2)}(\vec{k}_{1\perp}+\vec{k}_{2\perp}+
\vec{\lambda}_\perp-\vec{q}_\perp) \, , \nonumber
\end{eqnarray}
for the transverse-spin dependent case, and
\begin{eqnarray}
\frac{d\sigma^{uu}}
     {dy_1dy_2dP_\perp^2d^2\vec{q}_\perp}
&=&  \int d^2k_{1\perp}d^2k_{2\perp}d^2\lambda_\perp \,
x_aq_a^{\rm SIDIS}(x_a,k_{a\perp})\, x_bq_b^{\rm
SIDIS}(x_b,k_{b\perp})
\nonumber\\
&& 
\times \left[S_{qq'\to qq'}(\lambda_\perp)
      H_{qq'\to qq'}^{uu}(P_\perp^2)\right]_c \,
\delta^{(2)}(\vec{k}_{a\perp}+\vec{k}_{b\perp}
            +\lambda_\perp-\vec{q}_\perp) \; ,
\label{e2}
\end{eqnarray}
for the spin-averaged case. $H_{qq'\to qq'}$ and $S_{qq'\to qq'}$ are
partonic hard and soft factors, respectively, and the $[\quad ]_c$
represents a trace in color space between the hard and soft
factors due to the color flow into the jets
\cite{Botts:1989kf,Kidonakis:1997gm}. The hard factors in
Eqs.~(\ref{e4}),(\ref{e2}) only depend on the single hard scale
$P_\perp$ in terms of partonic Mandelstam variables of the
reaction $qq'\to qq'$, and
$x_a=\frac{P_\perp}{\sqrt{s}}\left(e^{y_1}+e^{y_2}\right)$,
$x_b=\frac{P_\perp}{\sqrt{s}}\left(e^{-y_1}+e^{-y_2}\right)$ with
$y_1$ and $y_2$ the rapidities of the two jets. In Eq.~(\ref{e4}),
$q_{T_a}^{\rm SIDIS}$ and $q_b^{\rm SIDIS}$ denote the
transverse-spin dependent Sivers quark distribution for hadron
$A$ and the unpolarized TMD quark distribution for hadron $B$,
respectively; these TMD parton distributions were chosen to follow
their definitions in the SIDIS process.

In the present model where the two initial partons have
separate charges, the associated parton distributions will depend
on these charges. For example, for a
polarized hadron with momentum $P_A=(P_A^+,0^-,0_\perp)$ with
$P_A^\pm=1/\sqrt{2}\left(P_A^0\pm P_A^3\right)$ and transverse
spin vector $\vec{S}_\perp$, the TMD distribution for quark flavor
$a$ with charge $g_1$ can be defined through the decomposition of
the following matrix element,
\begin{eqnarray}
{\cal M}_a &=&
\int\frac{P^+d\xi^-}{\pi}\,\frac{d^2\xi_\perp}{(2\pi)^2}\,
e^{-ix\xi^-P^++i\xi_\perp\cdot k_\perp} \langle
P_AS|\overline{\mit \psi}_a(\xi){\cal L}_{v_a}^\dagger(g_1;\infty;\xi)
          {\cal L}_{v_a}(g_1;\infty;0){\mit \psi}_a(0)|P_AS\rangle
\nonumber \\
&=&\frac{1}{2} \left[
 q_a^{\rm SIDIS}(x,k_\perp) \gamma_\mu P^\mu
+\frac{1}{M_P}
 q_{Ta}^{\rm SIDIS}(x,k_\perp)
 \epsilon_{\mu\nu\alpha\beta} \gamma^\mu P^\nu k^\alpha S^\beta
+ \dots \right] \ ,\label{e3}
\end{eqnarray}
where the gauge link ${\cal L}$ is defined in a covariant gauge as
\begin{equation}
{\cal L}_{{v_a}}(g_1;\infty;\xi) \equiv \exp\left(-ig_1\int^{\infty}_0
d\lambda \, {v_a}\cdot A(\lambda v +\xi)\right) \; ,
\label{glink}
\end{equation}
with the path extended to $+\infty$.  $v_a$ is a vector conjugate to the
momentum vector $P_A$. Since we will work in a covariant gauge
throughout this paper, the vector $v_a$ could be chosen to be a
light-cone vector with $v_a^2=0$ and $v_a\cdot P_A=1$. If we work
in a singular gauge, like the light-cone gauge, an additional
gauge link at spatial infinity ($\xi=+\infty$) will have to be
included in order to ensure the gauge invariance of the above
definitions \cite{BelJiYua02}.

Similarly, the quark distribution for the unpolarized hadron $B$
can be defined as
\begin{eqnarray}
{\cal M}_b &=&
\int\frac{P_B^-d\xi^+}{\pi}\,\frac{d^2\xi_\perp}{(2\pi)^2}\,
e^{-ix\xi^+P^-+i\xi_\perp\cdot k_\perp} \langle P_B|\overline{\mit
\psi}_b(\xi){\cal L}_{v_b}^\dagger(g_2;\infty;\xi)
          {\cal L}_{v_b}(g_2;\infty;0){\mit \psi}_b(0)|P_B\rangle
\nonumber \\
&=&\frac{1}{2} \left[
 q_b^{\rm SIDIS}(x,k_\perp) \gamma_\mu P^\mu
+ \dots \right] \ ,\label{e3p}
\end{eqnarray}
where the gauge link is ${\cal L}_{{v_b}}(g_2;\infty;\xi) \equiv
\exp\left(-ig_2\int^{\infty}_0 d\lambda \, {v_b}\cdot A(\lambda v
+\xi)\right)$, with $v_b$ a vector conjugate to $P_B$ satisfying $v_b^2=0$
and $v_b\cdot P_B=1$. Notice that in the above definition the
gauge link is associated with the coupling $g_2$.

At the lowest order, in the one-gluon contribution case, the relevant
hard factors in this model read:
\begin{eqnarray}
H_{qq'\to qq'}^{uu}&=&\frac{g_1^2g_2^2}{16\pi\hat s^2}
\frac{2(\hat s^2+\hat u^2)}{\hat t^2} \ , \label{hqqp} \nonumber\\
H_{qq'\to qq'}^{\rm Sivers}&=& \frac{g_1^2g_2^2}{16\pi\hat
s^2}\frac{g_1+2g_2}{g_1}\frac{2(\hat s^2+\hat u^2)}{\hat t^2} \ ,
\label{hqq}
\end{eqnarray}
where $\hat s,~\hat t,~\hat u$ are the partonic Mandelstam
variables. The above results can be directly obtained from
our previous result in \cite{qvy-long} by replacing there $C_u$ by $1$,
$C_I$ and $C_{F2}$ by $g_2/g_1$, and $C_{F1}$ by $1$. The expressions 
for the hard-scattering factors for other partonic channels may be found
in a similar way. We note that the above result for
$H_{qq'\to qq'}^{\rm Sivers}$ also agrees with that given
in~\cite{collinsqiu}.

Before we move on to the two-gluon exchange contributions,
we note that we have in fact also adapted our previous calculations
of~\cite{qvy-short,qvy-long} (see also~\cite{JiQiuVogYua06}) 
to this model case.
As in~\cite{qvy-short,qvy-long}, we have started
from a collinear-factorized framework and considered the one-gluon
contributions in the intermediate transverse momentum region
$\Lambda_{\rm QCD}\ll q_\perp\ll P_\perp$. We have found that
indeed the factorized structure of Eqs.~(\ref{e4}),(\ref{e2}) emerges,
with perturbatively defined TMD distributions that depend on the
charges $g_1$ and $g_2$ as given in Eqs.~(\ref{e3}),(\ref{e3p}),
and with the hard-scattering factors in~(\ref{hqq}). We refrain
from giving the details of these calculations in this paper, but
stress that the one-loop TMD factorization for the dijet-imbalance
found in~\cite{qvy-short,qvy-long} therefore extends
to the model of~\cite{collinsqiu} we consider here.

\begin{figure}[]
\begin{center}
\includegraphics[width=10cm]{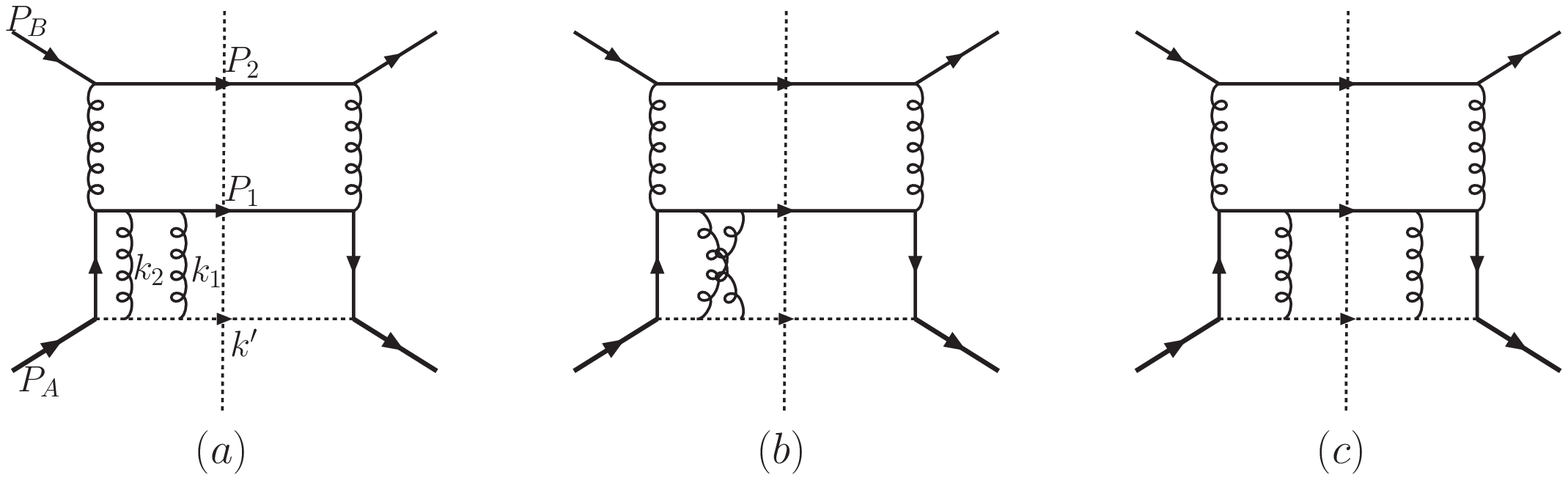}
\end{center}
\vskip -0.4cm \caption{\it Two-gluon exchange contributions
proportional to $g_1^2$. \label{fp1}}
\end{figure}

\begin{figure}[]
\begin{center}
\includegraphics[width=10cm]{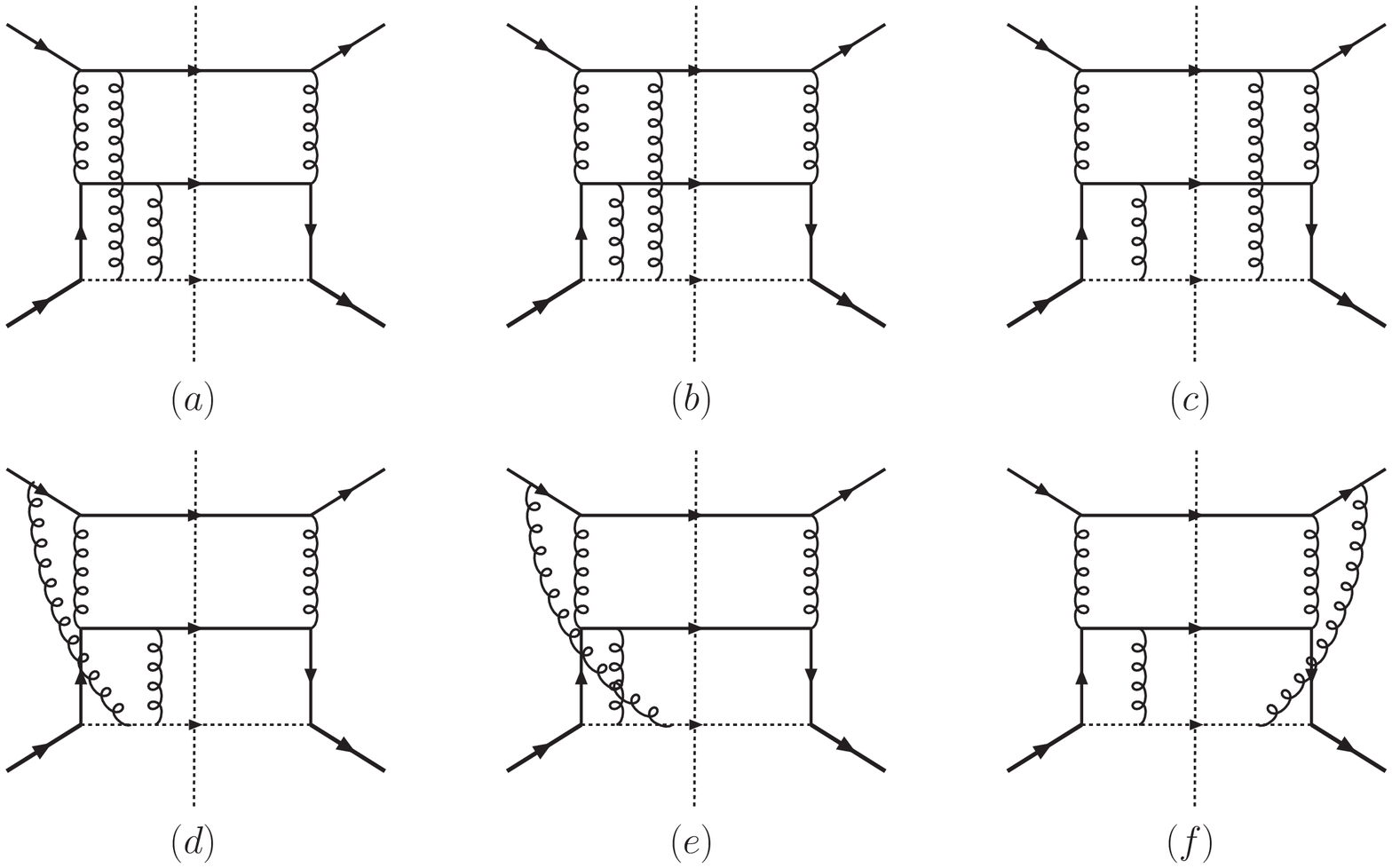}
\end{center}
\vskip -0.4cm \caption{\it Two-gluon exchange contributions
proportional to $g_1g_2$. \label{fp2}}
\end{figure}

\begin{figure}[]
\begin{center}
\includegraphics[width=10cm]{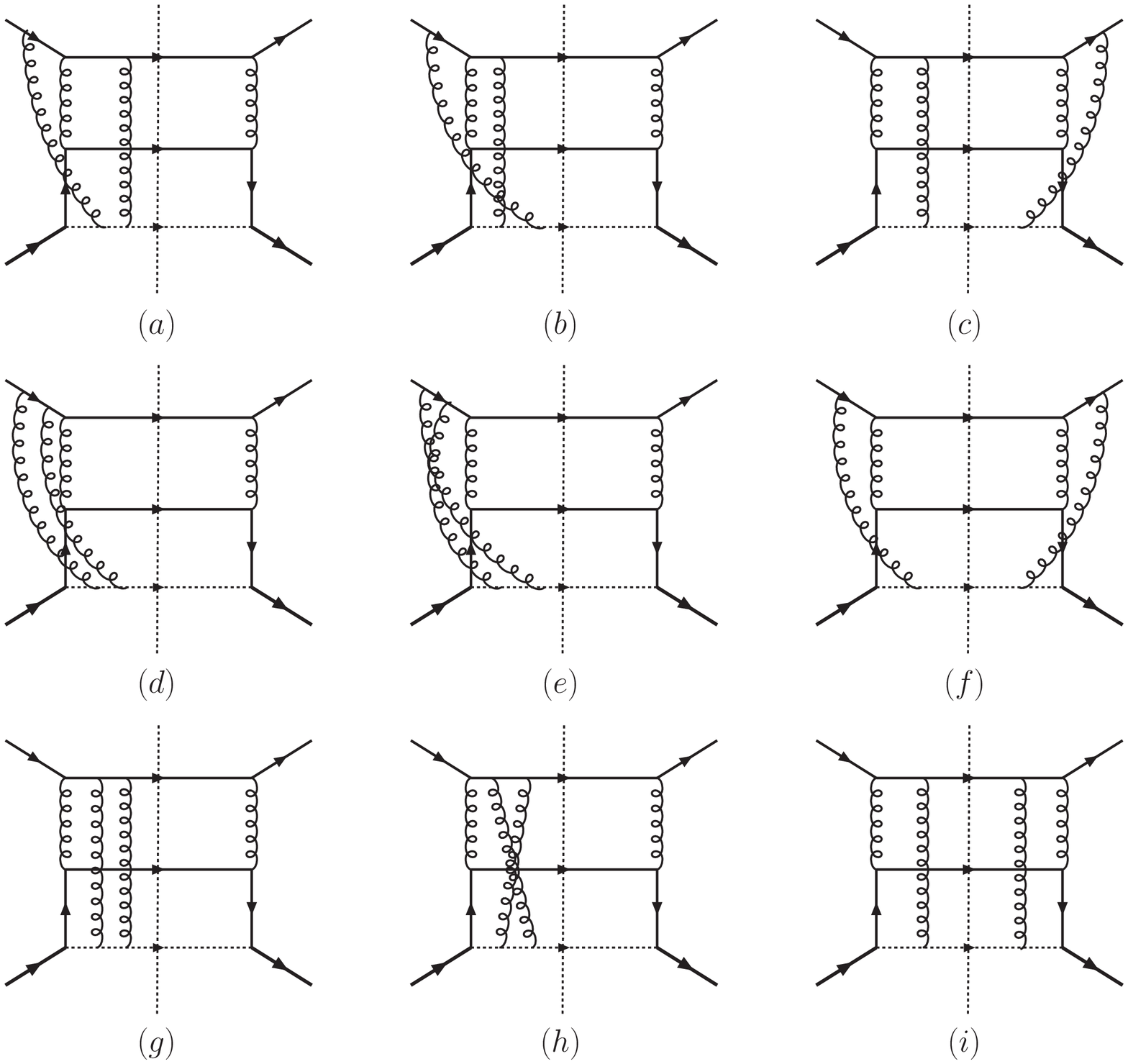}
\end{center}
\vskip -0.4cm \caption{\it Two-gluon exchange contributions
proportional to $g_2^2$. \label{fp3}}
\end{figure}

We now turn to the two-gluon exchange diagrams with the spectator of
hadron $A$, and their contributions to the SSA and the spin-averaged
cross section. They can be grouped into three parts: one proportional
to $g_1^2$ as shown in Fig.~\ref{fp1}, where both gluons attach to
the outgoing quark line with charge $g_1$; one proportional to $g_1g_2$
as shown in Fig.~\ref{fp2}, where one gluon attaches to $g_1$ quark line and
another gluon to the incoming or outgoing $g_2$ quark line; and one
proportional to $g_2^2$ as shown in Fig.~\ref{fp3}, where both
gluons attach to the $g_2$ quark line.


Let us first discuss the contributions from the diagrams in
Fig.~\ref{fp1}. In the following calculations, we only consider
the dominant contributions from the exchange of gluons
collinear to the polarized hadron $A$, which survive at the leading
power of $q_\perp/P_\perp$. We can then utilize the
eikonal approximation~\cite{qvy-long,mulders,mulders1,collinsqiu}. The contribution
from the diagram in Fig.~\ref{fp1}(a) will then be proportional to
\begin{eqnarray}
&&g_1^2\int
\frac{d^4k_1}{(2\pi)^4}\frac{d^4k_2}{(2\pi)^4}\frac{1}
{(P_A-k'-k_1-k_2)^2-m^2+i\epsilon}
\frac{1}{(k'+k_1)^2-\mu^2+i\epsilon}\frac{1}{(k'+k_1+k_2)^2-
\mu^2+i\epsilon}\nonumber\\
&&~~~~~~~~~~~~~~~~~~~\times\frac{1}{k_1^2-\lambda^2+i\epsilon}
\frac{1}{k_2^2-\lambda^2+i\epsilon}
\left(\frac{1}{-k_1^++i\epsilon}\right)\left(\frac{1}{-k_1^+-k_2^+
+i\epsilon}\right)\ , \label{eik1}
\end{eqnarray}
where $k_1$ and $k_2$ are the momenta of the two exchanged gluons, 
and where we have
introduced the masses $m,\mu,\lambda$ for the scattering quarks
and for the gluon, respectively. The last two factors in~(\ref{eik1})
come from the eikonal approximations for the two gluon attachments
to the outgoing quark line $g_1$. A similar contribution is obtained
from diagram (b), except for a difference in the eikonal
propagators, for which we now have
\begin{equation}
\left(\frac{1}{-k_2^++i\epsilon}\right)\left(\frac{1}{-k_1^+-k_2^+
+i\epsilon}\right)\
.
\end{equation}
Thus, the total contribution by these two diagrams will depend
on the following expression for the eikonal propagators,
\begin{equation}
A=g_1^2\left[\left(\frac{1}{-k_1^++i\epsilon}\right)
\left(\frac{1}{-k_1^+-k_2^++i\epsilon}\right)+
\left(\frac{1}{-k_2^++i\epsilon}\right)\left(
\frac{1}{-k_1^+-k_2^++i\epsilon}\right)\right]
\ ,
\end{equation}
and the rest of the expression is identical for these two diagrams.
The imaginary part of the above expression contributes
to the SSA, whereas the real part contributes to the unpolarized
cross section. We can further simplify the above term as
\begin{equation}
A=g_1^2\left[\frac{1}{k_1^+k_2^+}+(-i\pi)^2\delta(k_1^+)\delta
(k_2^+)\right] +i\pi
g_1^2\left[\frac{\delta(k_1^+)}{k_2^+}+\frac{\delta(k_2^+)}{k_1^+}\right]
\ .
\end{equation}

Similar calculations can be performed for the diagrams in
Figs.~\ref{fp2} and~\ref{fp3}. The contributions from diagrams
2 (a,b,d,e) will be proportional to the factor
\begin{eqnarray}
B&=&g_1g_2\left[2\left(\frac{1}{-k_1^++i\epsilon}\right)
\left(\frac{1}{-k_2^++i\epsilon}\right)+
\left(\frac{1}{-k_1^++i\epsilon}\right)\left(\frac{1}{k_2^+
+i\epsilon}\right)\right.\nonumber\\
&&~~~~~~~~\left.+
\left(\frac{1}{k_1^++i\epsilon}\right)\left(\frac{1}{-k_2^+
+i\epsilon}\right)\right]
\nonumber\\
&=&g_1g_2\left[4(-i\pi)^2\delta(k_1^+)\delta (k_2^+)\right] +i2\pi
g_1g_2\left[\frac{\delta(k_1^+)}{k_2^+}+\frac{\delta(k_2^+)}{k_1^+}\right]
\ ,
\end{eqnarray}
where partial cancellations between the diagrams have occurred
in the final result. The contributions from diagrams
3 (a,b,d,e,g,h) will depend on the factor
\begin{eqnarray}
C&=&g_2^2\left[
\left(\frac{1}{-k_1^++i\epsilon}\right)\left(\frac{1}{k_2^++i\epsilon}\right)
+
\left(\frac{1}{k_1^++i\epsilon}\right)\left(\frac{1}{-k_2^++i\epsilon}\right)
\right.\nonumber\\
&&~~~~~+
\left(\frac{1}{k_1^++i\epsilon}\right)\left(\frac{1}{k_1^++k_2^+
+i\epsilon}\right)+
\left(\frac{1}{k_2^++i\epsilon}\right)\left(\frac{1}{k_1^++k_2^+
+i\epsilon}\right)\nonumber\\
&&~~~~~\left.+
\left(\frac{1}{-k_1^++i\epsilon}\right)\left(\frac{1}{-k_1^+-k_2^+
+i\epsilon}\right)+
\left(\frac{1}{-k_2^++i\epsilon}\right)\left(\frac{1}{-k_1^+-k_2^+
+i\epsilon}\right)
\right]
\nonumber\\
&=&g_2^2\left[4(-i\pi)^2\delta(k_1^+)\delta (k_2^+)\right] \ .
\end{eqnarray}
Again, there are cancellations between the diagrams, which in this case
leave us with only a real part contribution.

The total contribution from the diagrams discussed above thus becomes
\begin{eqnarray}
A+B+C&=&g_1^2\left[\frac{1}{k_1^+k_2^+}+(-i\pi)^2\delta(k_1^+)
\delta(k_2^+)\right]+
g_1(g_1+2g_2)
(i\pi)\left[\frac{\delta(k_2^+)}{k_1^+}+\frac{\delta(k_1^+)}{k_2^+}
\right]\nonumber\\[3mm]
&&+4\left(g_1g_2+g_2^2\right)(-i\pi)^2\delta(k_1^+)\delta(k_2^+) \
.\label{abc}
\end{eqnarray}
As we mentioned above, the first and third terms will contribute
to the unpolarized cross section, whereas the second term will
contribute to the SSA. Clearly, the first term can be factorized
into the unpolarized TMD quark distribution, multiplied by the
leading-order hard factor $H_{qq'\to qq'}^{uu}$ in
Eq.~(\ref{hqqp}). The second term can also be factorized into the
quark Sivers function for hadron $A$, multiplied by the
leading-order hard factor $H_{qq'\to qq'}^{\rm Sivers}$ in
Eq.~(\ref{hqq}). The third term, however, cannot be factorized
into the unpolarized quark distribution defined in Eq.~(\ref{e3}),
multiplied by the leading hard factor $H_{qq'\to qq'}^{uu}$.

In order to account for this uncancelled contribution, we have to modify
the gauge link definition in Eq.~(\ref{glink}):
\begin{eqnarray}
{\cal L}_{v_a}(g_1;\infty;\xi)\to {\cal L}'_{v_a}(g_1,g_2;\xi)&\equiv&
{\cal P}\exp\left(-ig_1\int_0^\infty d\lambda v_a\cdot A(\xi+\lambda
v_a)\right)\nonumber\\
&&\times{\cal P}\exp\left(-ig_2\int_0^\infty d\lambda v_a\cdot
A(\xi+\lambda v_a)\right)\nonumber\\
&&\times{\cal P}\exp\left(ig_2\int_0^{-\infty} d\lambda v_a\cdot
A(\xi+\lambda v_a)\right) \ ,\label{g3}
\end{eqnarray}
which corresponds to what has been proposed in \cite{mulders,mulders1}. The
first term takes into account the contributions due to final-state
interactions of the outgoing quark with charge $g_1$, the second
one those of the outgoing quark with charge $g_2$, and the third
one the initial-state interactions of the incoming quark with
charge $g_2$. With this modification of the gauge link in the
definition of the TMD quark distributions, the above results 
can all be reproduced by the two-gluon exchange diagram contributions, 
multiplied by the leading order hard factor. This is true for both
the unpolarized and the spin-dependent (Sivers) case, even though
for the latter, as seen from Eq.~(\ref{abc}),  
the complications that require the redefinition of 
the gauge link are not yet visible at this order but should first
occur at the next order of perturbation theory. 

A similar analysis can be performed for the diagrams in
Fig.~\ref{fp1}(c), Fig.~\ref{fp2}(c,f), and Fig.~\ref{fp3}(c,f,i).
All their contributions are reproduced by the above
modified TMD quark distribution, multiplied by the leading-order
hard factor. It is important to note that these diagrams do not
contribute to the SSA. Moreover, their contributions, along
with that in the last term of 
Eq.~({\ref{abc}), will lead to an infrared-finite
contribution to the differential unpolarized cross section. It
will be interesting to investigate further how this
will affect the dijet-correlation at hadron colliders.

In the QED-like model of Ref.~\cite{collinsqiu}, 
further study shows that we can
simplify the gauge link in Eq.~(\ref{g3}) to
\begin{eqnarray}
{\cal L}'_{v_a}(\xi)&=& {\cal P}\exp\left(-ig_1\int_0^\infty
d\lambda v_a\cdot A(\xi+\lambda
v_a)\right)\nonumber\\
&&\times{\cal P}\exp\left(-ig_2\int_{-\infty}^\infty d\lambda
v_a\cdot A(\xi_\perp+\lambda v_a)\right)\ .
\end{eqnarray}
We notice that the last factor will reduce to a pure phase when
integrated over the transverse momentum, because in that case
$\xi_\perp$ is set to zero and there is no other dependence on
$\xi$. In this case, a collinear factorization approach is
appropriate, and the gauge link will be defined in a standard way
\cite{Collins:1989gx}. In particular, the last term of
Eq.~(\ref{abc}) will be cancelled out by diagrams (c) and (c,f)
of Figs.~\ref{fp2} and \ref{fp3}, respectively.

{\bf 4. Conclusions.} In this paper, we have compared three
recent theoretical studies \cite{mulders,mulders1,qvy-short,qvy-long,collinsqiu}
of the factorization properties of dijet-correlations in a polarized 
hadronic reaction when the two jets are produced nearly back-to-back 
in azimuthal angle. We have shown that the first-order perturbative results
of~\cite{qvy-short,qvy-long} are reproduced when the gauge links
derived in~\cite{mulders,mulders1} are expanded to that order. Within the
abelian model considered in~\cite{mulders,collinsqiu}, we have also
calculated the two-gluon contributions to the initial-state
and final-state interactions and verified that the results are
consistent with the expected gauge-link structure. We therefore conclude
that the results of the various studies are mutually consistent. They
show that beyond the first order, the TMD quark distributions need to be
modified in a process-dependent way, in order for them to correctly take
into account collinear gluon contributions. This is a departure from standard
factorization, which on the other hand still does not exclude that the dijet
observable obeys a more generalized factorization in terms of more
complicated TMD correlators. Further studies will be needed here.

\section*{Acknowledgments}
We are grateful to Mauro Anselmino, Alessandro Bacchetta,
Daniel Boer, Cedran Bomhof, John Collins, Markus Diehl, Andreas Metz, 
Piet Mulders, Jianwei Qiu, and Peter
Schweitzer for useful discussions. We thank
RIKEN, Brookhaven National Laboratory and the
U.S. Department of Energy (contract number DE-AC02-98CH10886) for
providing the facilities essential for the completion of their
work.



\begin{thebibliography}
\frenchspacing

\bibitem{BoeVog03}
  D.~Boer and W.~Vogelsang,
  Phys.\ Rev.\ D {\bf 69}, 094025 (2004).


\bibitem{mulders} C.~J.~Bomhof, P.~J.~Mulders and F.~Pijlman,
  Phys.\ Lett.\ B {\bf 596}, 277 (2004).

\bibitem{mulders1} C.~J.~Bomhof, P.~J.~Mulders and F.~Pijlman,
  Eur.\ Phys.\ J.\ C {\bf 47}, 147 (2006);
A.~Bacchetta, C.~J.~Bomhof, P.~J.~Mulders and F.~Pijlman,
  Phys.\ Rev.\ D {\bf 72}, 034030 (2005);
  C.~J.~Bomhof and P.~J.~Mulders,
  JHEP {\bf 0702}, 029 (2007).

\bibitem{VogYua05}
  W.~Vogelsang and F.~Yuan,
  Phys.\ Rev.\ D {\bf 72}, 054028 (2005).

\bibitem{dijet-cor1}
  C.~J.~Bomhof, P.~J.~Mulders, W.~Vogelsang and F.~Yuan,
  Phys.\ Rev.\  D {\bf 75}, 074019 (2007).

\bibitem{qvy-short}
  J.~W.~Qiu, W.~Vogelsang and F.~Yuan,
  arXiv:0704.1153 [hep-ph], Phys. Lett. {\bf B650}, 373 (2007).

\bibitem{qvy-long}
  J.~W.~Qiu, W.~Vogelsang and F.~Yuan,
  arXiv:0706.1196 [hep-ph].


\bibitem{collinsqiu} 
  J.~Collins and J.~W.~Qiu,
  arXiv:0705.2141 [hep-ph].


\bibitem{RT}
  P.~G.~Ratcliffe and O.~V.~Teryaev,
  arXiv:hep-ph/0703293.


\bibitem{star-dijet1}
  B.~I.~Abelev, {\it et al.}, [STAR Collaboration]
  arXiv:0705.4629 [hep-ex];
 J.~Balewski, talk presented at the SPIN 2006 Symposium,
Kyoto, Japan, October 2-7, 2006, arXiv:hep-ex/0612036.


\bibitem{Siv90}
D.~W.~Sivers,
Phys.\ Rev.\ D {\bf 41}, 83 (1990);
Phys.\ Rev.\ D {\bf 43}, 261 (1991).


\bibitem{BroHwaSch02}
S.~J.~Brodsky, D.~S.~Hwang and I.~Schmidt,
Phys.\ Lett.\ B {\bf 530}, 99 (2002);
Nucl.\ Phys.\ B {\bf 642}, 344 (2002).

\bibitem{Col02}
J.~C.~Collins,
Phys.\ Lett.\ B {\bf 536}, 43 (2002).


\bibitem{BelJiYua02}
X.~Ji and F.~Yuan,
Phys.\ Lett.\ B {\bf 543}, 66 (2002);
A.~V.~Belitsky, X.~Ji and F.~Yuan,
Nucl.\ Phys.\ B {\bf 656}, 165 (2003).

\bibitem{Boer:2003cm}
  D.~Boer, P.~J.~Mulders and F.~Pijlman,
  Nucl.\ Phys.\  B {\bf 667}, 201 (2003)
  [arXiv:hep-ph/0303034].


\bibitem{et}
  A.~V.~Efremov and O.~V.~Teryaev,
  Sov.\ J.\ Nucl.\ Phys.\  {\bf 36}, 140 (1982)
  [Yad.\ Fiz.\  {\bf 36}, 242 (1982)];
  A.~V.~Efremov and O.~V.~Teryaev,
  Phys.\ Lett.\ B {\bf 150}, 383 (1985).

\bibitem{qiusterman}
J.~Qiu and G.~Sterman,
Phys.\ Rev.\ Lett.\  {\bf 67}, 2264 (1991);
  Nucl.\ Phys.\ B {\bf 378}, 52 (1992);
Phys.\ Rev.\ D {\bf 59}, 014004 (1999).


\bibitem{Botts:1989kf}
  J.~Botts and G.~Sterman,
  Nucl.\ Phys.\  B {\bf 325}, 62 (1989).

\bibitem{Kidonakis:1997gm}
  N.~Kidonakis and G.~Sterman,
  Nucl.\ Phys.\  B {\bf 505}, 321 (1997);
Nucl.\ Phys.\ B {\bf 525}, 299 (1998);
 N.~Kidonakis, G.~Oderda and G.~Sterman,
Nucl.\ Phys.\  B {\bf 531}, 365 (1998).


\bibitem{JiQiuVogYua06}
  X.~Ji, J.~W.~Qiu, W.~Vogelsang and F.~Yuan,
Phys.\ Rev.\ Lett.\ {\bf 97}, 082002 (2006);
  Phys.\ Rev.\ D {\bf 73}, 094017 (2006);
  Phys.\ Lett.\ B {\bf 638}, 178 (2006).

\bibitem{Collins:1989gx}
  J.~C.~Collins, D.~E.~Soper and G.~Sterman,
  Adv.\ Ser.\ Direct.\ High Energy Phys.\  {\bf 5}, 1 (1988)
  [arXiv:hep-ph/0409313].




\end{thebibliography}
\end{document}